\documentclass[a4paper,11pt,openany]{article}
\usepackage{authblk}
\usepackage{lineno}
\usepackage{mathrsfs}
\usepackage{geometry}
\usepackage{amssymb}
\usepackage{amsmath}
\usepackage{booktabs}
\usepackage{enumerate}
\usepackage{verbatim}
\usepackage{graphicx}
\usepackage{epsfig}
\usepackage{epsf}
\usepackage{epstopdf}
\usepackage{amstext}
\usepackage{latexsym}
\usepackage{subfigure}
\usepackage[matrix,frame,arrow]{xypic}

\begin{document}
\title{Effect of Temperature and Charged Particle Fluence on the
Resistivity 
of Polycrystalline CVD
Diamond Sensors}



\renewcommand\Authands{, } 
\renewcommand\Affilfont{\itshape\small} 

\author{Rui Wang}
\author{Martin Hoeferkamp}
\author{Sally Seidel}


\affil{Department of Physics and Astronomy, University of New Mexico, Albuquerque, NM 87108, USA}

\date{\today}


\maketitle

\begin{abstract}
The resistivity of
polycrystalline chemical vapor deposition
diamond sensors is studied in samples exposed to
fluences relevant to the environment of the High Luminosity Large Hadron Collider.
We measure the leakage
current
for a range of bias voltages on samples irradiated with 800 MeV
protons up to $1.6\times 10^{16}$p/cm$^{2}$.
The proton beam at LANSCE, Los Alamos National Laboratory, was applied to
irradiate the samples. The devices' resistivity is extracted for
temperatures in the $-10^\circ$C to $+20^\circ$C range.
\end{abstract}

\section{Introduction}
Polycrystalline chemical vapor deposition (CVD) diamond~\cite{marinelli,sellin,schloegl,brambilla} is presently used in particle physics experiments for beam
monitoring~\cite{bcm,cms}, and it is
being further developed~\cite{modules} for use in vertexing and tracking detectors planned for the challenging
radiation environment of the inner layers of High Luminosity Large Hadron Collider (HL-LHC)
detectors.  At
distances shorter than about 24~cm from an LHC collision (i.e., the regime covered
by inner pixel detectors), the dominant source of radiation damage is charged particles.
Any effect of radiation damage upon the resistivity of the detection material will, if
uncompensated, propagate to
the leakage current; the result would be that all
assessments of the material properties that depend upon leakage current measurement,
including active volume and charge collection distance, would be impacted.
Accordingly we have
studied the resistivity
of polycrystalline CVD diamonds as a function of
temperature and
proton fluence.  The 800 MeV proton beam at LANSCE, Los Alamos, was used in the irradiation.    This
study was carried out in the framework of
the RD42 Collaboration at CERN.

\section{Method}
The devices (see Table~\ref{tab1}), manufactured by Element Six~\cite{manu}, are polycrystalline
and structured with metalized
pads and backplane. The cleaning and metalization process is based on a technique in~\cite{Zhao}.
That process begins with application of three heavily oxidizing acids to remove all organic residues
and leave the surface oxygen terminated.  The sequence is HCl-HNO$_3$ (3:1), H$_2$SO$_4$ (3:2), then
H$_2$SO$_4$-H$_2$O$_2$ (1:1).  This is followed by an oxygen plasma etch for 4 minutes.  After the
high energy sputter by composite
TiW, the contacts are annealed for another 4 minutes at $450^\circ$C in an inert atmosphere.

The device thicknesses were measured with an Eichhorn and Hausmann Contactless Wafer
Thickness and Geometry Gauge (model MX 203-6-33) and confirmed optically with a microscope; their lengths and
widths were measured optically.  These diamonds
are taken from the same series, number 1006115, produced in 2008.
As the results from the two devices are consistent, for clarity
the graphs
in this paper display the 1006115-36 data unless otherwise indicated (these are indicated in the
legends of the graphs as ``15-36.'')  Device
1006115-36 was exposed to fluences $3.85 \times 10^{15}$, $1.11 \times 10^{16}$, $1.36 \times 10^{16}$, and
$1.63 \times 10^{16}$ p/cm$^2$.  Device 1006115-46 (indicated in the graphs as ``15-46'')
was exposed to fluences $2.76 \times 10^{15}$ and $7.5 \times 10^{15}$
p/cm$^2$.

\begin{table}
    \begin{center}
    \begin{tabular}{|c|c|}
      \hline
      Diamond sensor & Dimensions (cm$\times$cm$\times\mu$m) \\
      \hline
      1006115-36 & $1.016 \times 1.017 \times 440$  \\
      1006115-46 & $1.007 \times 1.008 \times 432$ \\

      \hline
    \end{tabular}
\caption{Dimensions and features of the diamond sensors used in this study.
Representative uncertainties on these lengths are 0.002 cm on the transverse
dimensions and 10~$\mu$m on the thickness.}
\label{tab1}
    \end{center}
\end{table}

\begin{figure}[htbp]
        \centering
\subfigure[Measurement configuration 1.]{
\includegraphics[width=10cm]{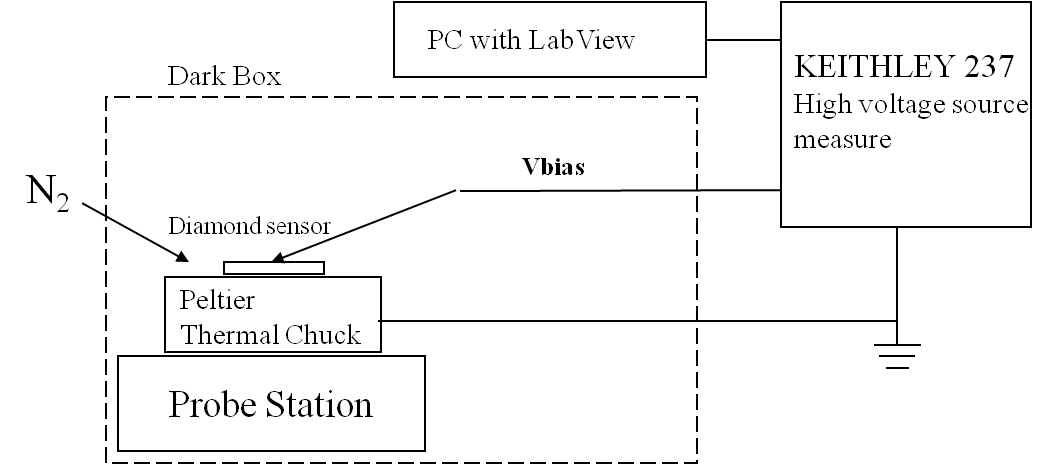}
\label{setup:15-36}}
\subfigure[Measurement configuration 2.]{
\includegraphics[width=14cm]{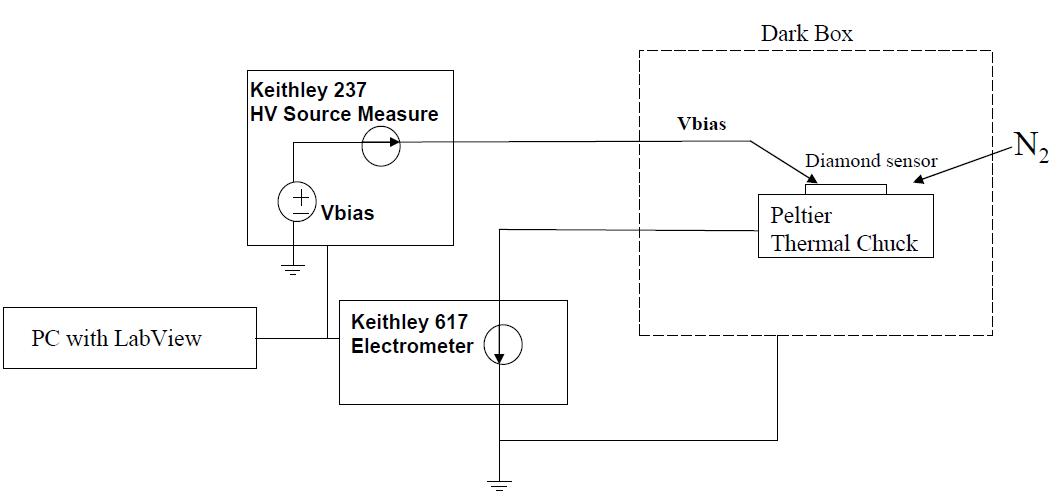}
\label{setup:15-46}}
        \caption{The experimental setups for measuring leakage current as a
function of bias voltage.}
\label{fig1}
        \end{figure}

Resistivity $\rho$ is computed as
\begin{equation}
\label{eq:resistivity}
\rho=AR/d,
\end{equation}
where $R$ is the inverse of the slope of a
linear fit to a graph of leakage current versus bias
voltage (``IV''); $A$ is the area of the sensor under test;
and $d$ is the sensor thickness.
Two slightly different setups, see Figure~\ref{fig1}, were used for the IV measurement in order
to quantify systematic uncertainty associated with the instrument configuration.
With ground applied to the detector back side and high voltage applied to the
front,
the bulk leakage current data are acquired by the Keithley 237 source measure
unit in Setup 1 and by the Keithley 617 electrometer in Setup 2.  The advantage of Setup 1 is its simplicity:
a single instrument is used to measure the sourced current.  The advantage of Setup 2 is the fact that
the measurement is made instead on the returned current (so sourced current that did not cross the
bulk is excluded); the disadvantage is that this setup requires two instruments, each with its own intrinsic
contribution to measurement uncertainty.
Dry N$_2$ is
applied continuously to the environment to prevent condensation.
The sensor temperature is
maintained at approximately $-10^\circ$C, $0^\circ$C, $10^\circ$C or $20^\circ$C
by the thermal chuck on which the sensor rests. Relative
humidity is less than 5\% for all measurements below
$20^\circ$ and less than 35\% for room temperature measurements.
Bias voltage is ramped over
the range from -500 V to +500 V with confirmation measurements in both
directions.  Data taken with positive and negative voltage are fitted separately
for voltages with magnitude of 200 V and higher.  (We exclude data for voltages in the realm of
100 V, as these currents are comparable to the intrinsic accuracy of the Keithley devices
which is 100 fA.)
The
separate fits are consistent, and the
slope $R$ is their average.

\begin{figure}[htbp]
        \centering
        \includegraphics[width=12cm]{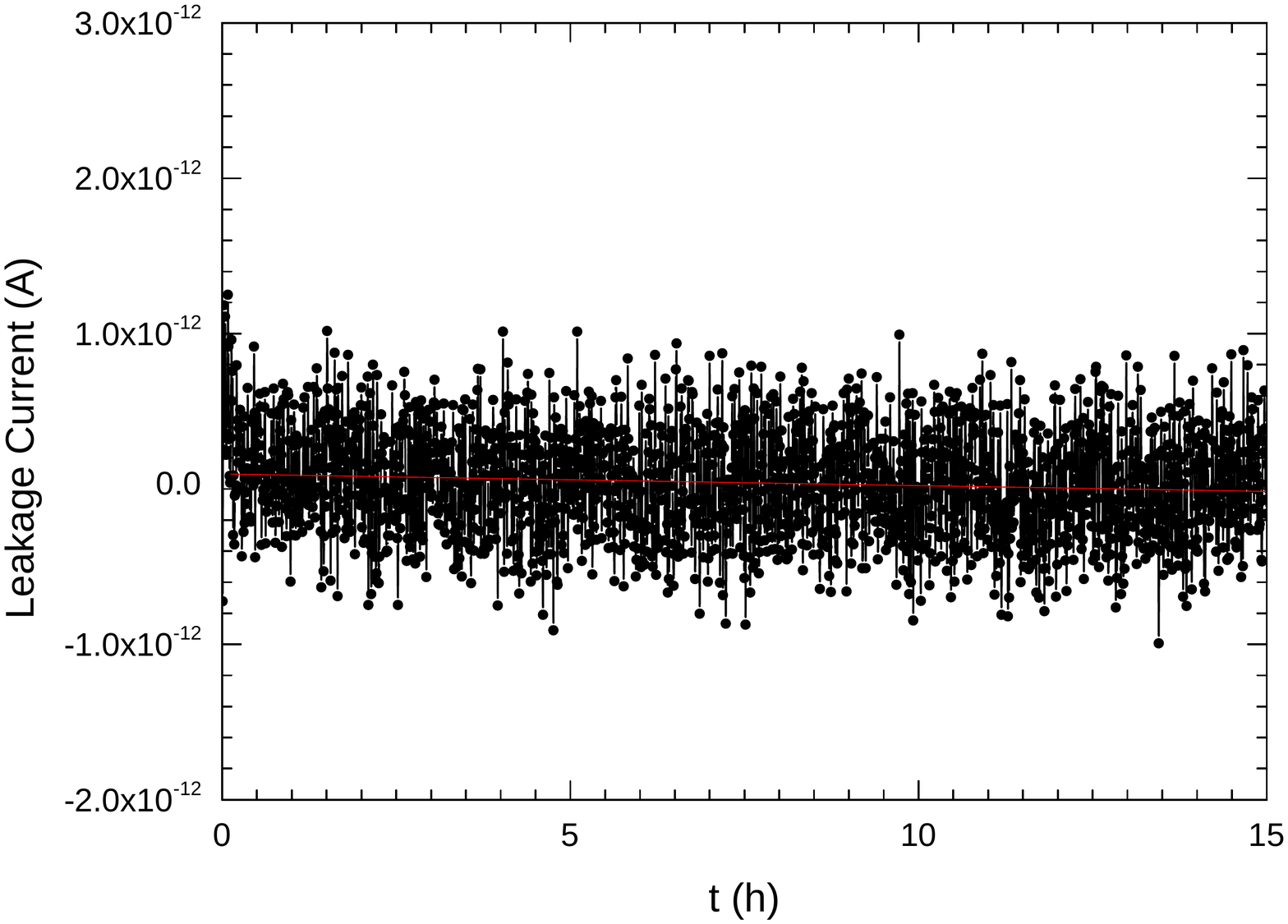}
        \caption{This current versus time graph for a typical measurement illustrates the stability
of the current.  Measurements
commence about an hour after the change in bias voltage and are recorded every 30 seconds thereafter.
These data were taken at $20^\circ$C
on device 1006115-36 after
it had received a fluence of $1.63 \times 10^{16}$ p/cm$^2$.}
\label{fig:stability}
        \end{figure}

The standard deviation on any measured current varies between 3 and 9
$\times 10^{-13}$~A, depending upon the setup, derived from
three to five measurements under identical conditions.  The standard deviation
is unaffected by the temperature or humidity within the ranges used here.
An interval of a few hours is
typically allocated to the measurement of a single IV point. After being
mounted to the thermal chuck, the diamond's current and temperature are
monitored continuously. The temperature of the diamond is recorded through
a thermal sensor mounted directly to the thermal chuck. Temperature uncertainty
is less than $0.1^\circ$ C for any individual measurement and falls
in the range $0.2^\circ$ to $0.8^\circ$ C for the full voltage scan of most
devices.  Equilibration takes approximately 30 minutes.
An average current and temperature are extracted from data during the period
beginning about one hour after installation and continuing up to four hours.
We do not observe any deviations of the average slope from flatness
during these intervals.
Figure~\ref{fig:stability} demonstrates the stability of the current at a
typical bias point and illustrates the size of a standard deviation on
any measured current.  That particular measurement involved application of 500 V over 15 hours
at $20^\circ$ C to diamond 1006115-36 after it had received $1.63 \times 10^{16}$ p/cm$^2$.
The line fitted to the graph for all data after 30 minutes intercepts current $90 \pm 17$ fA
with slope of $-5.74 \times 10^{-17} \pm 1.61 \times 10^{-17}$ A/hr, i.e., consistent with zero.
For the interval from 1 hr to 4 hr over which a measurement is taken, the slope of the data
is $-2.29 \times 10^{-16} \pm 1.65 \times 10^{-16}$ A/hr.

\section{Results}

    \subsection{Leakage Current}

\begin{figure}[htbp]
\centering
\subfigure[Unirradiated sensor.]{
\includegraphics[width=7cm]{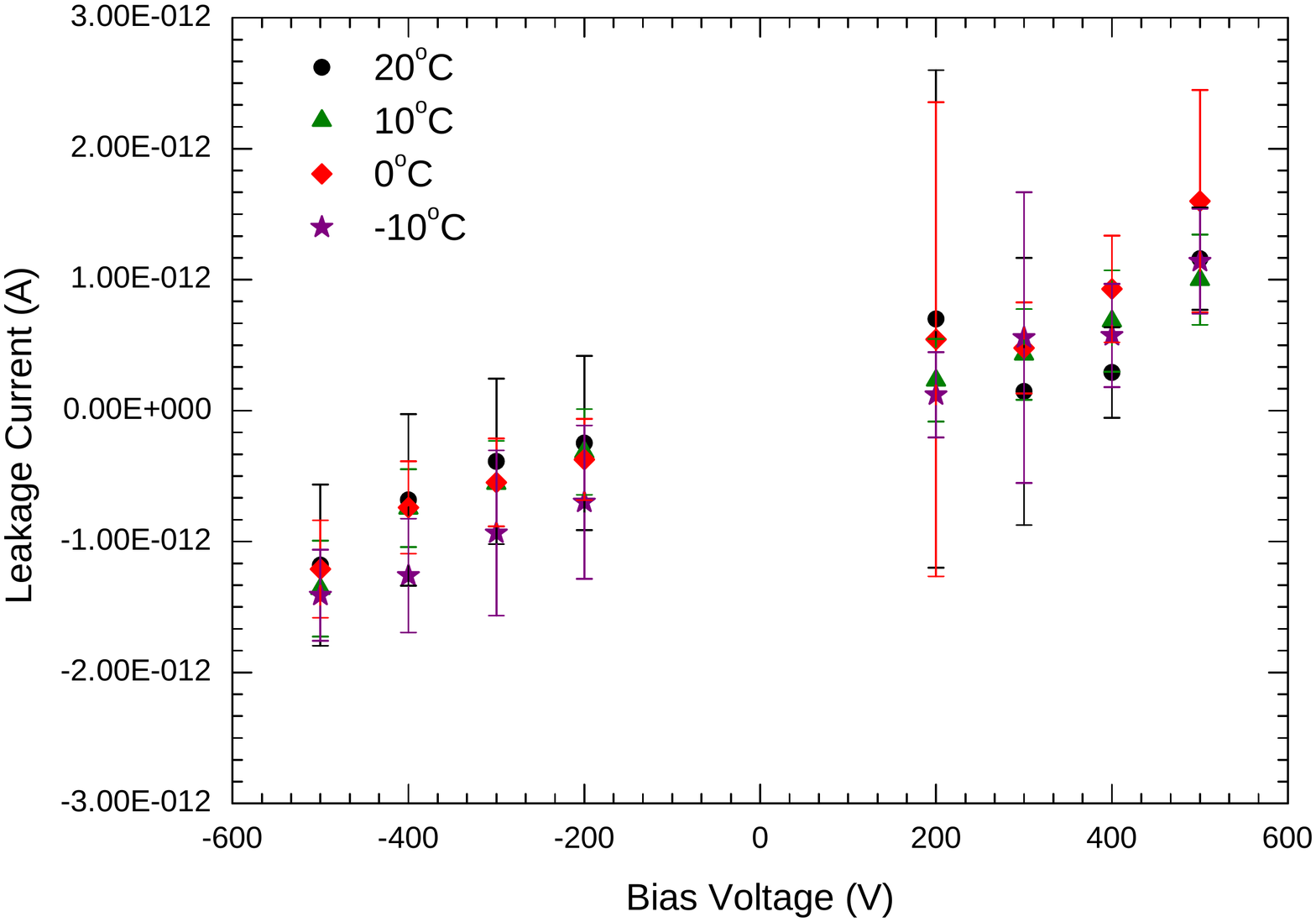}
\label{fig:IV_0}}
\subfigure[Sensor irradiated to $3.85 \times 10^{15}$ 800 MeV p cm$^{-2}$.]{
\includegraphics[width=7cm]{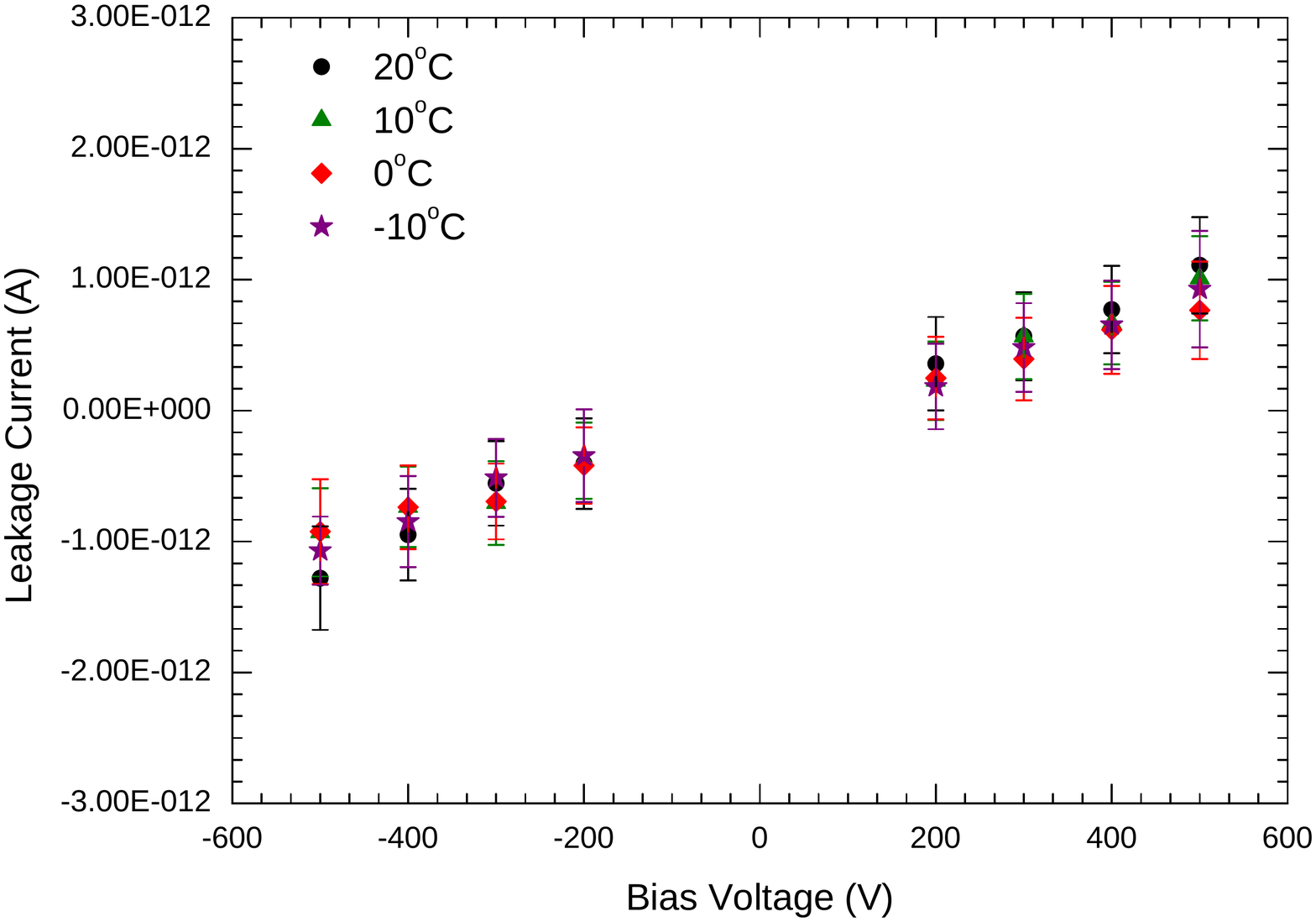}
\label{fig:IV_3.85E15}}
\subfigure[Sensor irradiated to $1.11 \times 10^{16}$ 800 MeV p cm$^{-2}$.]{
\includegraphics[width=7cm]{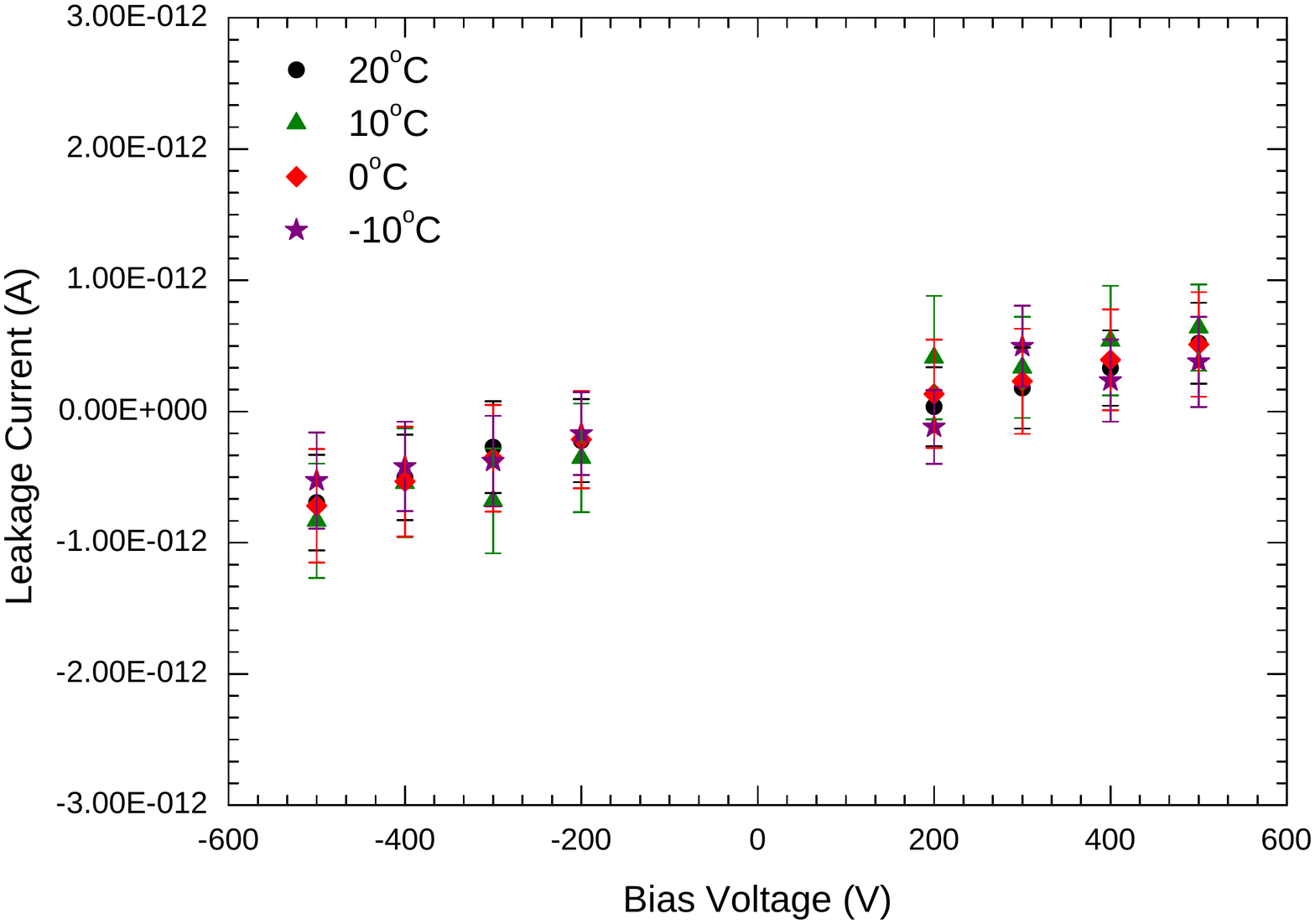}
\label{fig:IV_1.11E16}}
\subfigure[Sensor irradiated to $1.36 \times 10^{16}$ 800 MeV p cm$^{-2}$.]{
\includegraphics[width=7cm]{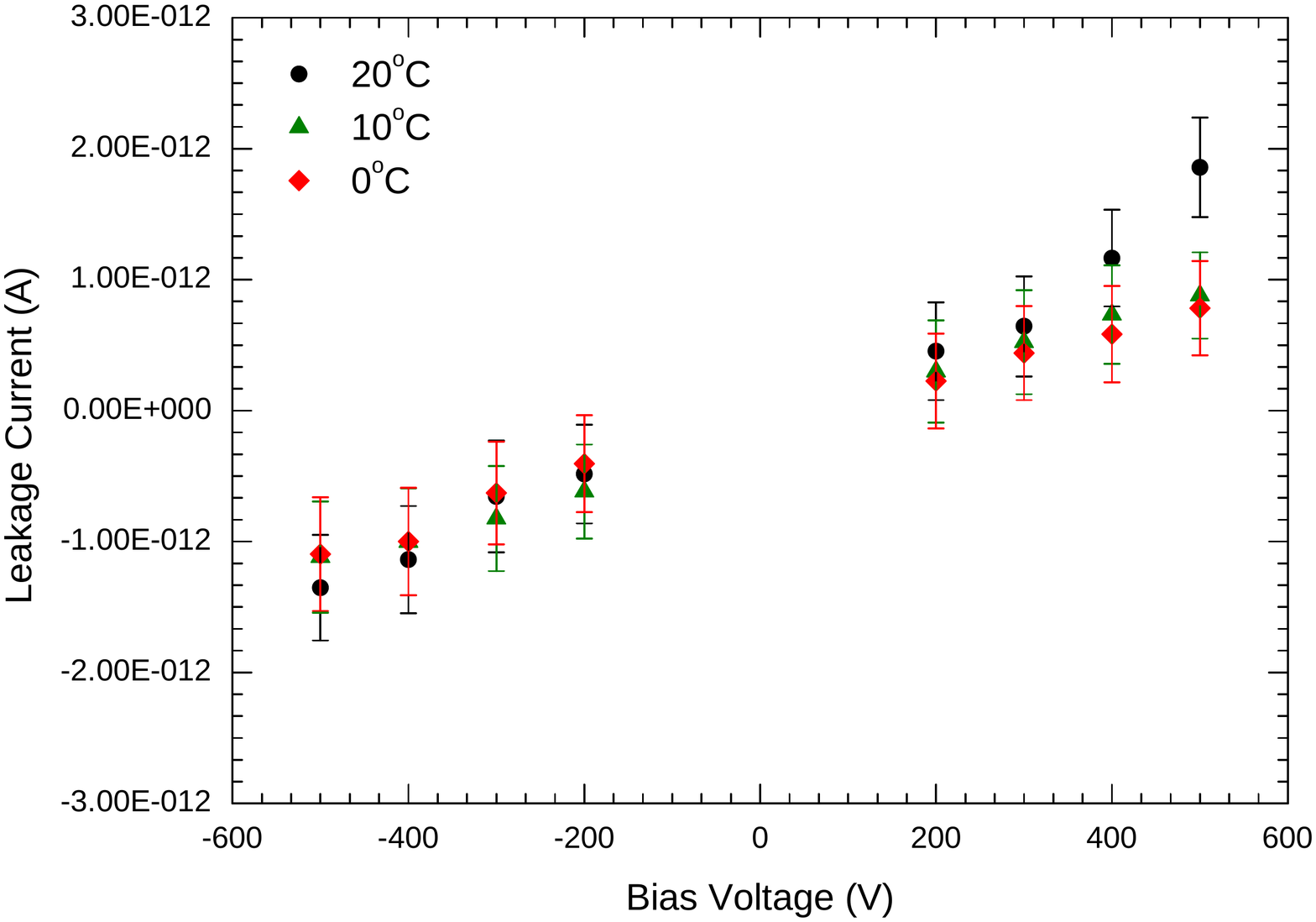}
\label{fig:IV_1.36E16}}
\subfigure[Sensor irradiated to $1.63 \times 10^{16}$ 800 MeV p cm$^{-2}$.]{
\includegraphics[width=7cm]{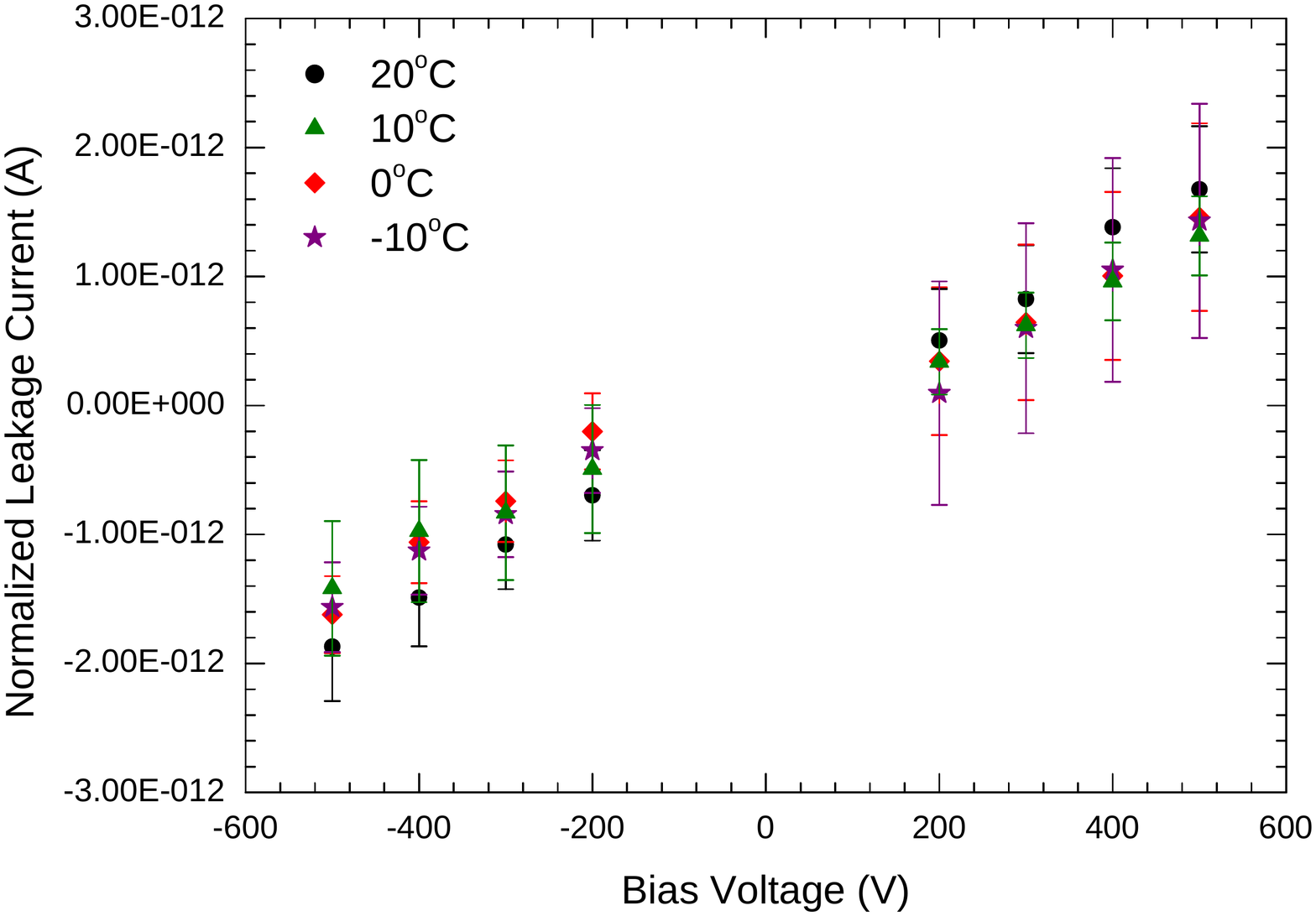}
\label{fig:IV_1.63E16}}
\caption{Leakage current versus bias voltage in a diamond sensor irradiated to 5 fluence
levels and measured at 4 temperatures.}
\label{fig:5ivs}
\end{figure}

Figure~\ref{fig:5ivs}
shows the leakage current in diamond 1006115-36
for positive and negative bias voltages up to
500 V.   This diamond showed breakdown just above 500 V prior to irradiation, and
that set the scale for our studies.
After receiving fluences of $3.85 \times 10^{15}$ and $1.11 \times 10^{16}$
p/cm$^2$, however, it was tested to 1000 V without breakdown.  At fluence
$1.63 \times 10^{16}$ p/cm$^2$, it again displayed
breakdown just above 500 V.  Diamond 1006115-46 similarly showed breakdown near 500 V
before irradiation.  After irradiation it was tested up to 800 V without breakdown.

\begin{figure}[htbp]
\centering
\subfigure[Unirradiated sensor]{
\includegraphics[width=12cm]{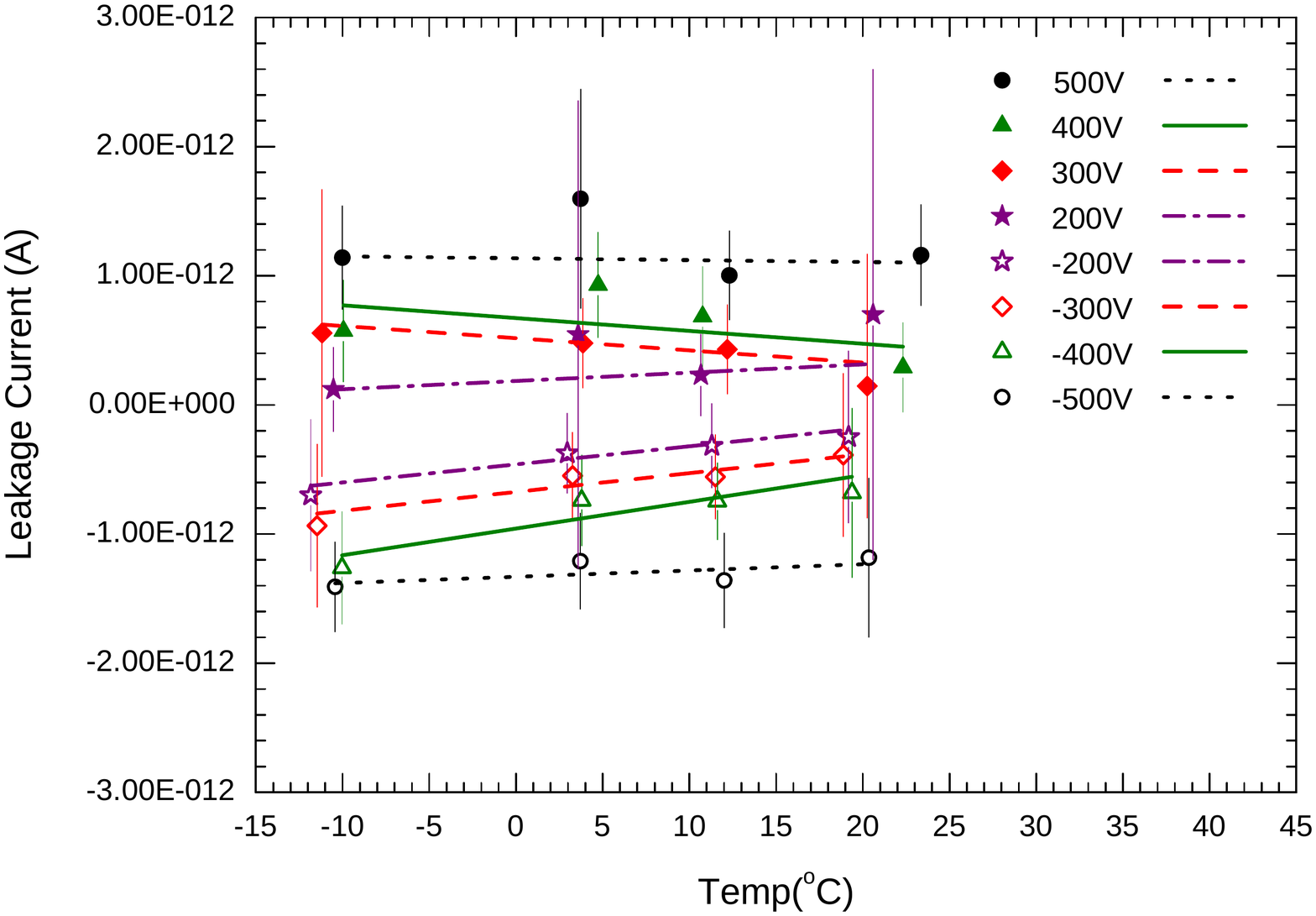}
\label{fig:IvsT1}}
\subfigure[Sensor irradiated to $1.1\times10^{16}$ 800 MeV p cm$^{-2}$]{
\includegraphics[width=12cm]{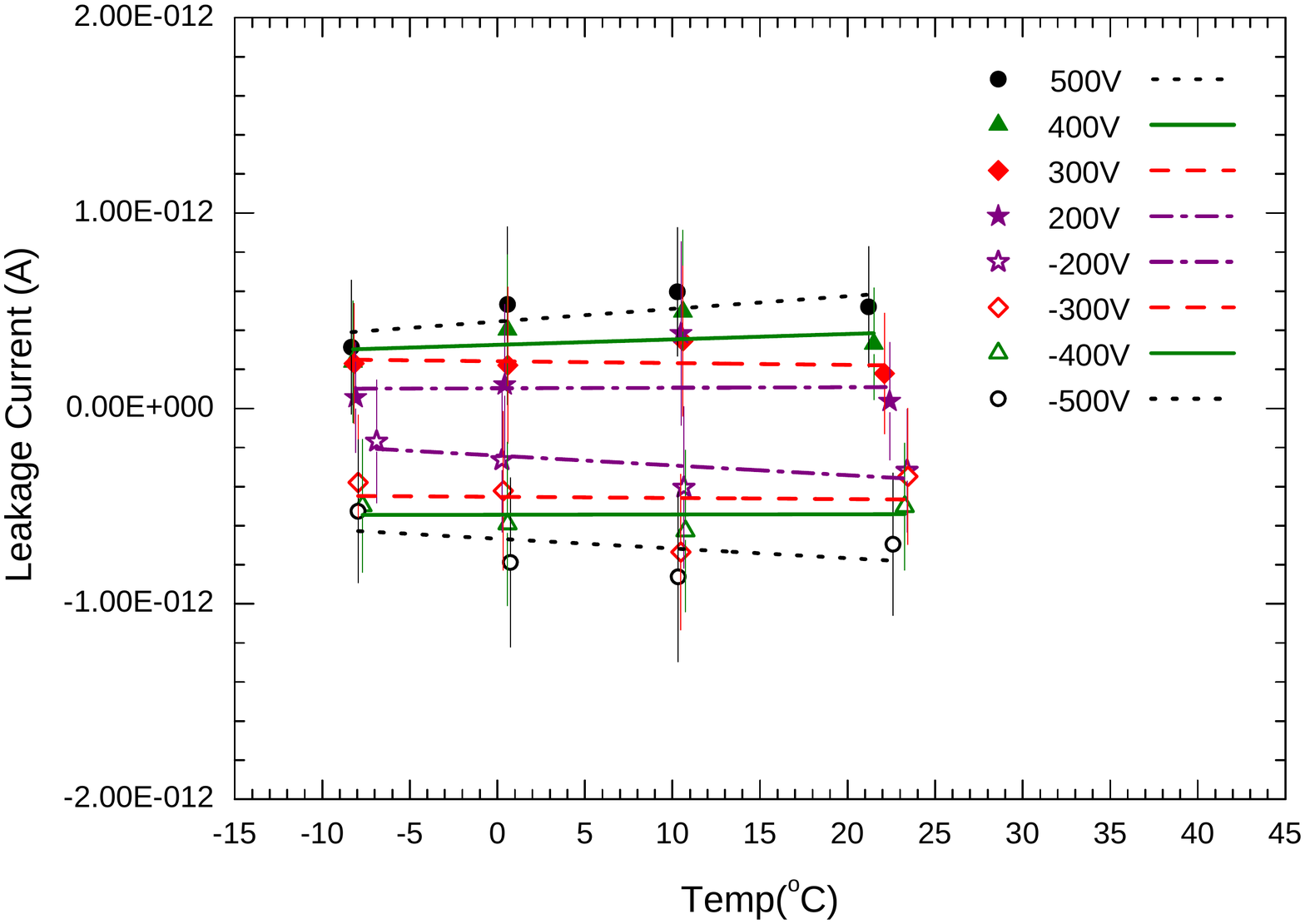}
\label{fig:IvsT2}}
\caption{Leakage current in (a) an unirradiated diamond sensor and (b) a
sensor that has been irradiated to $1.11 \times 10^{16}$ 800 MeV p cm$^{-2}$, for positive
and negative bias voltages up to 500 V, with linear fits of unconstrained slope to each dataset.  The range of values of $\chi^2$/dof for these fits is 0.1 to 1.1.}
\label{fig:ivst}
\end{figure}

\begin{figure}[htbp]
\centering
\subfigure[Unirradiated sensor]{
\includegraphics[width=12cm]{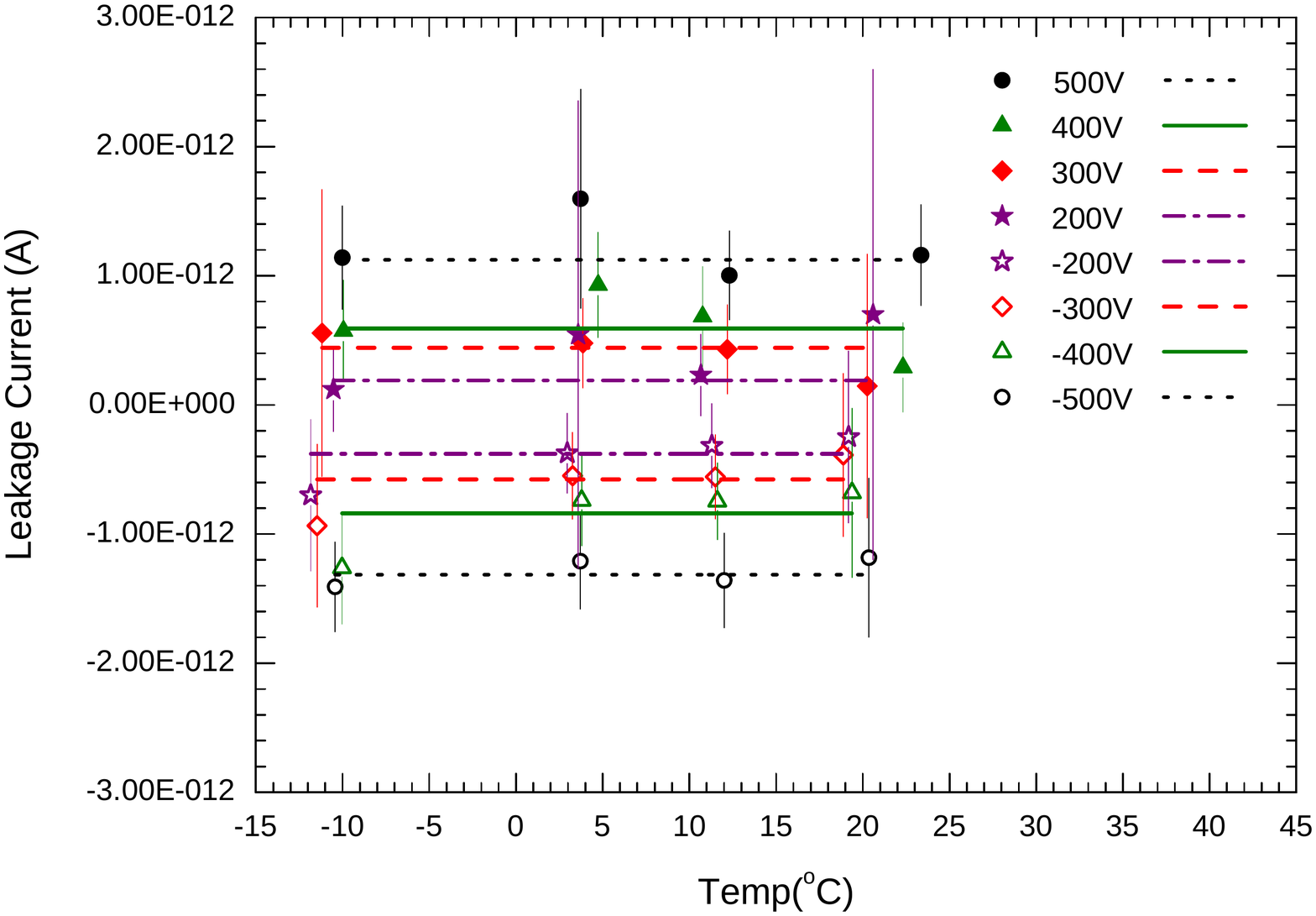}
\label{fig:IvsT1fixed}}
\subfigure[Sensor irradiated to $1.1\times10^{16}$ 800 MeV p cm$^{-2}$]{
\includegraphics[width=12cm]{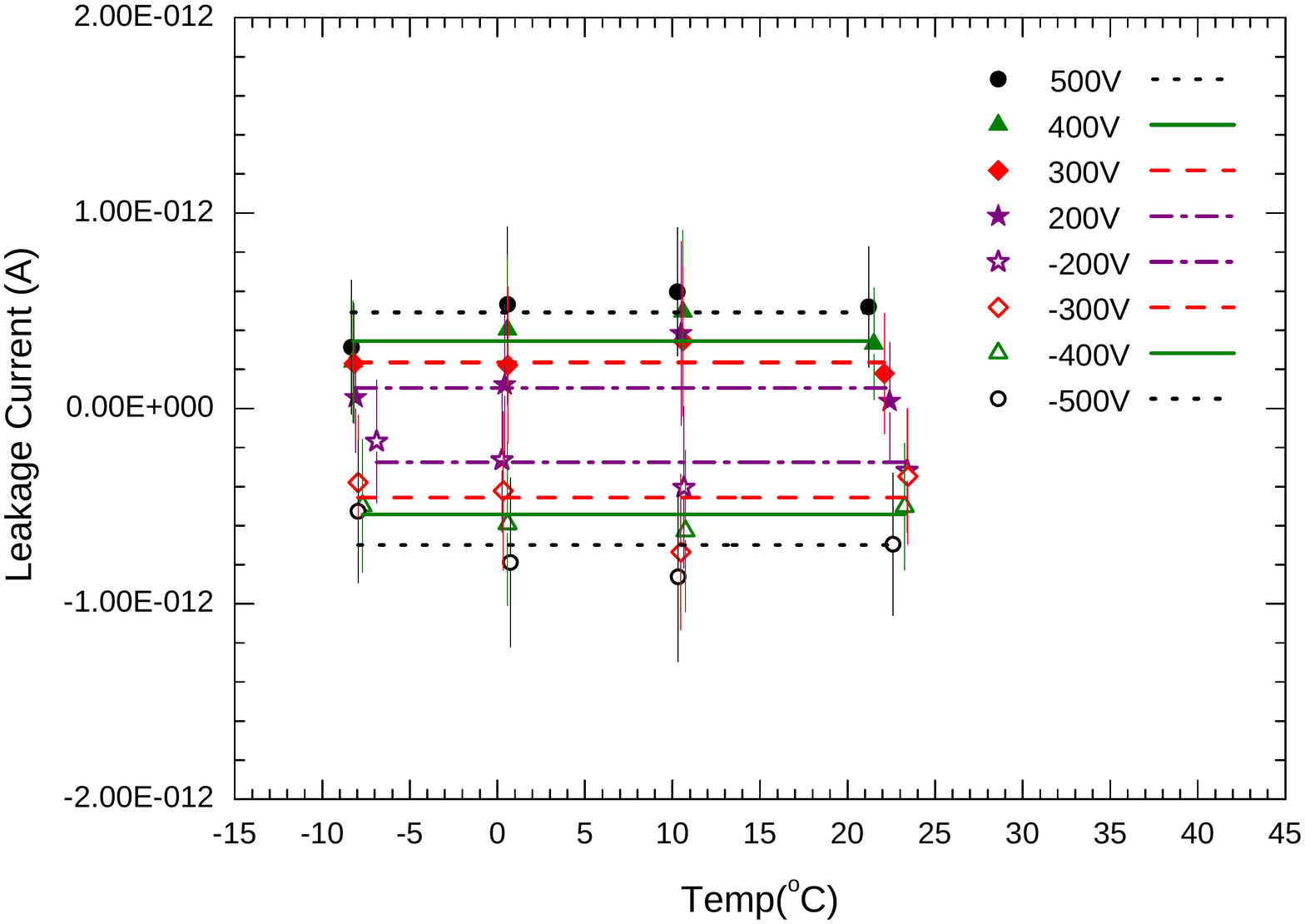}
\label{fig:IvsT2fixed}}
\caption{Leakage current in (a) an unirradiated diamond sensor and (b) a
sensor that has been irradiated to $1.11 \times 10^{16}$ 800 MeV p cm$^{-2}$, for positive
and negative bias voltages up to 500 V, with linear fits for slope fixed to zero to each dataset.  The range of values of $\chi^2$/dof for these fits is 0.1 to 1.5.}
\label{fig:ivstfixed}
\end{figure}

The IV characteristics are consistent for temperatures
in the range $-10^\circ$ to $20^\circ$ C for fluences up to $1.63 \times 10^{16}$ p/cm$^2$. (An instrument failure caused the data taken at $-10^\circ$C after the
$1.36 \times 10^{16}$ exposure to be lost.)
We observe no dependence of leakage current upon temperature in this range:
Figure~\ref{fig:ivst} displays
this information for two example fluences with straight-line fits, slope unconstrained, to the data.  The values of
$\chi^2$/dof for the fits in Figure~\ref{fig:ivst} range from 0.1 to 1.1.  If the fits are
redone for slopes fixed to zero (Figure~\ref{fig:ivstfixed}), the range of $\chi^2$/dof values is 0.1 to 1.5.

    \subsection{Resistivity}

\begin{figure}[htbp]
        \centering
        \includegraphics[width=13cm]{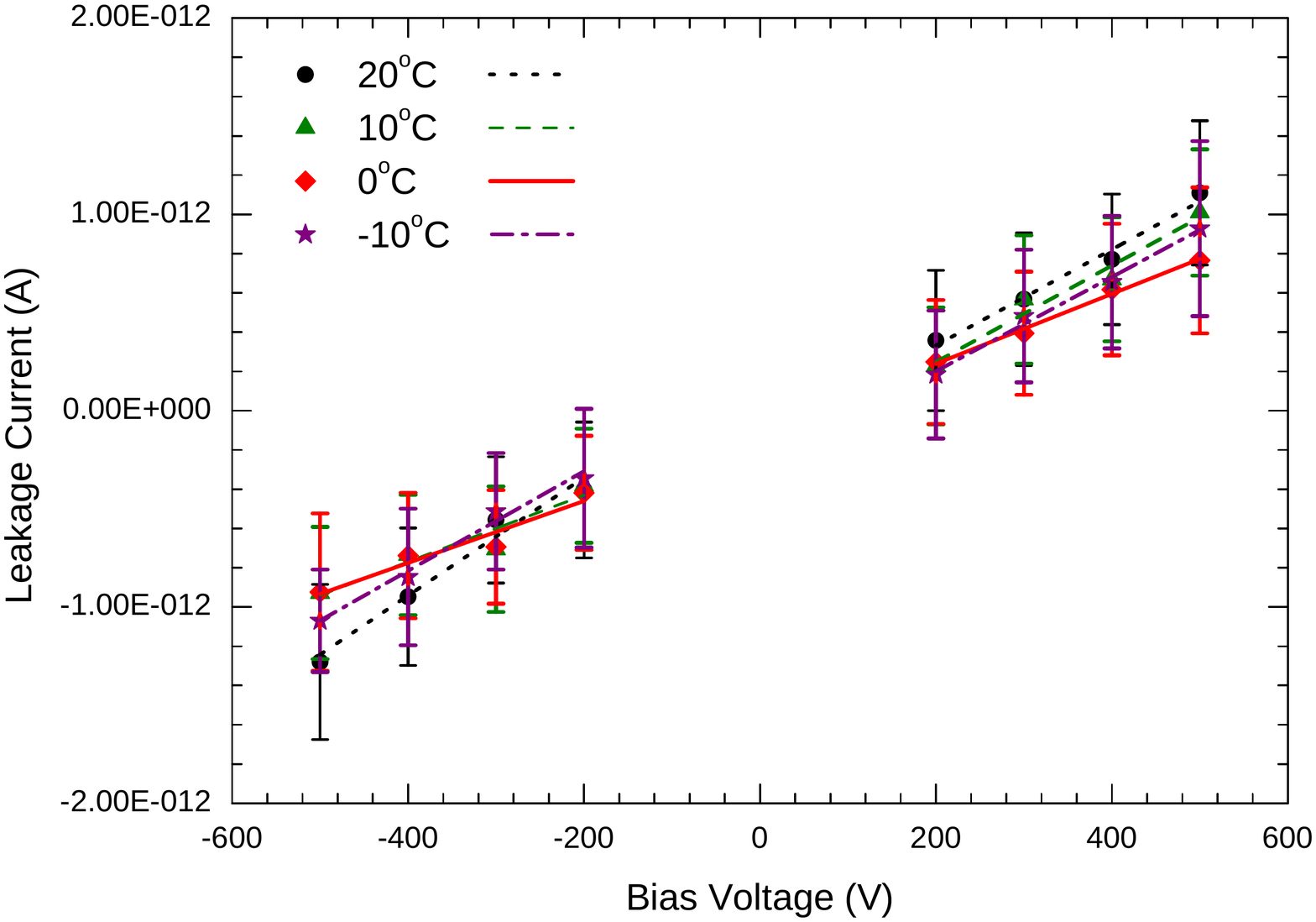}
        \caption{Leakage current as a function of bias voltage in an
example sensor measured at 4 temperatures after exposure to a fluence of
$3.85 \times 10^{15}$ 800 MeV protons/cm$^2$, with fitted lines
superimposed.}
\label{diamondIV-ir.pdf}
       \end{figure}

The data in each IV graph are fitted to straight
lines for the two separate ranges [-500 V, -200 V] and [200 V, 500 V].  For each temperature
and fluence combination, those two slopes are extracted and
averaged,
and this average $R$ is converted to a resistivity using
Equation~\ref{eq:resistivity}.  To illustrate the method,
Figure~\ref{diamondIV-ir.pdf} shows the set of fitted lines
resulting from this procedure applied to sensor 1006115-36 after exposure to
$3.85 \times 10^{15}$ 800 MeV protons/cm$^2$.

\begin{figure}[htbp]
          \centering
       \includegraphics[width=12cm]{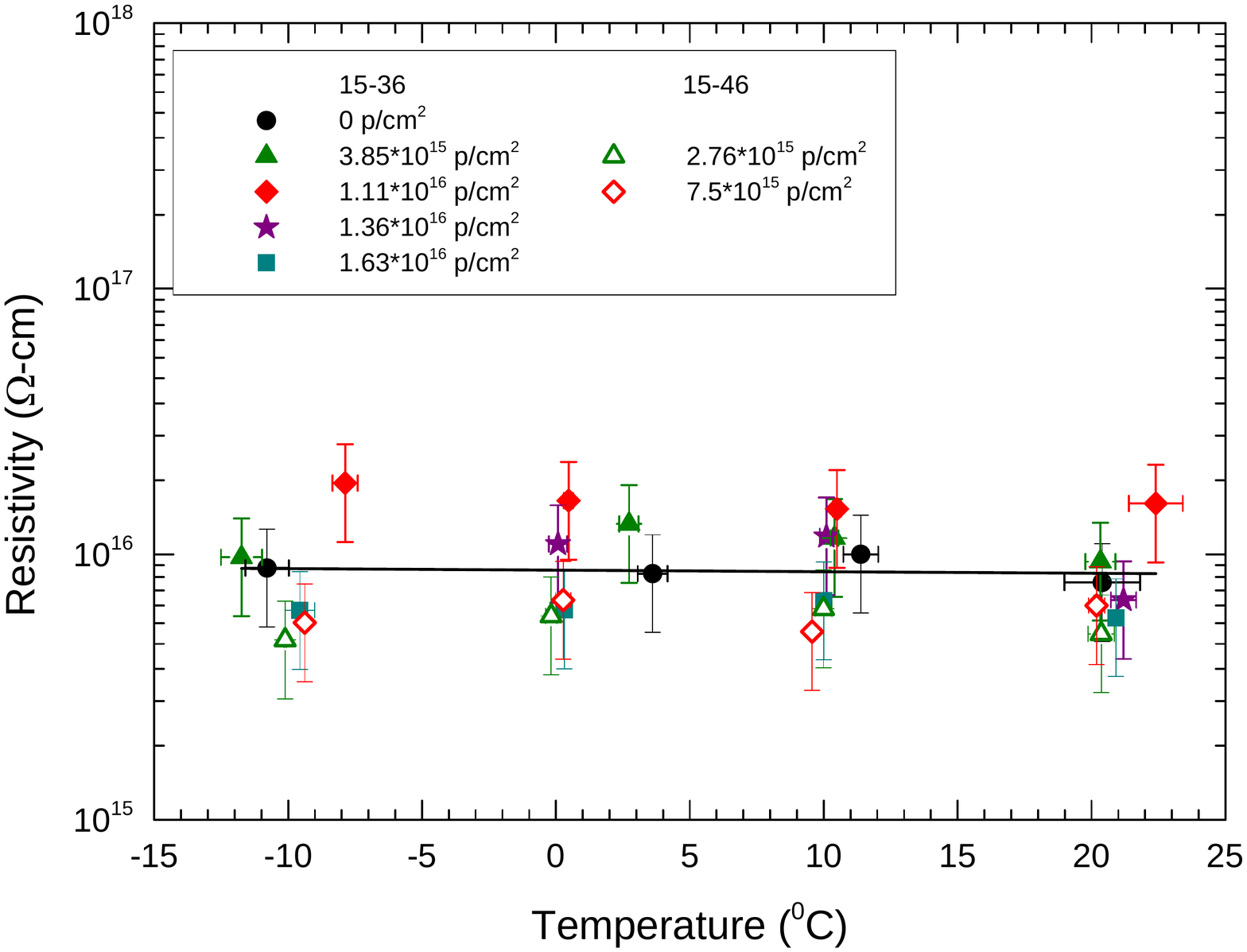}
        \caption{Resistivity of the diamond sensors as a function of temperature, for fluences ranging from
 0 to $1.63 \times 10^{16}$ 800 MeV protons/cm$^{2}$.  A free linear fit to the data is shown, with intercept
 $(8.37 \pm 0.55) \times 10^{15} \Omega$-cm, slope $(-0.63 \pm 4.13) \times 10^{13} \Omega$-cm/$^\circ$C, and $\chi^2$/dof 0.62.  The
error bars show the combination of statistical and systematic uncertainty.}
        \label{diamondrt}
      \end{figure}

\begin{figure}[htbp]
          \centering
       \includegraphics[width=12cm]{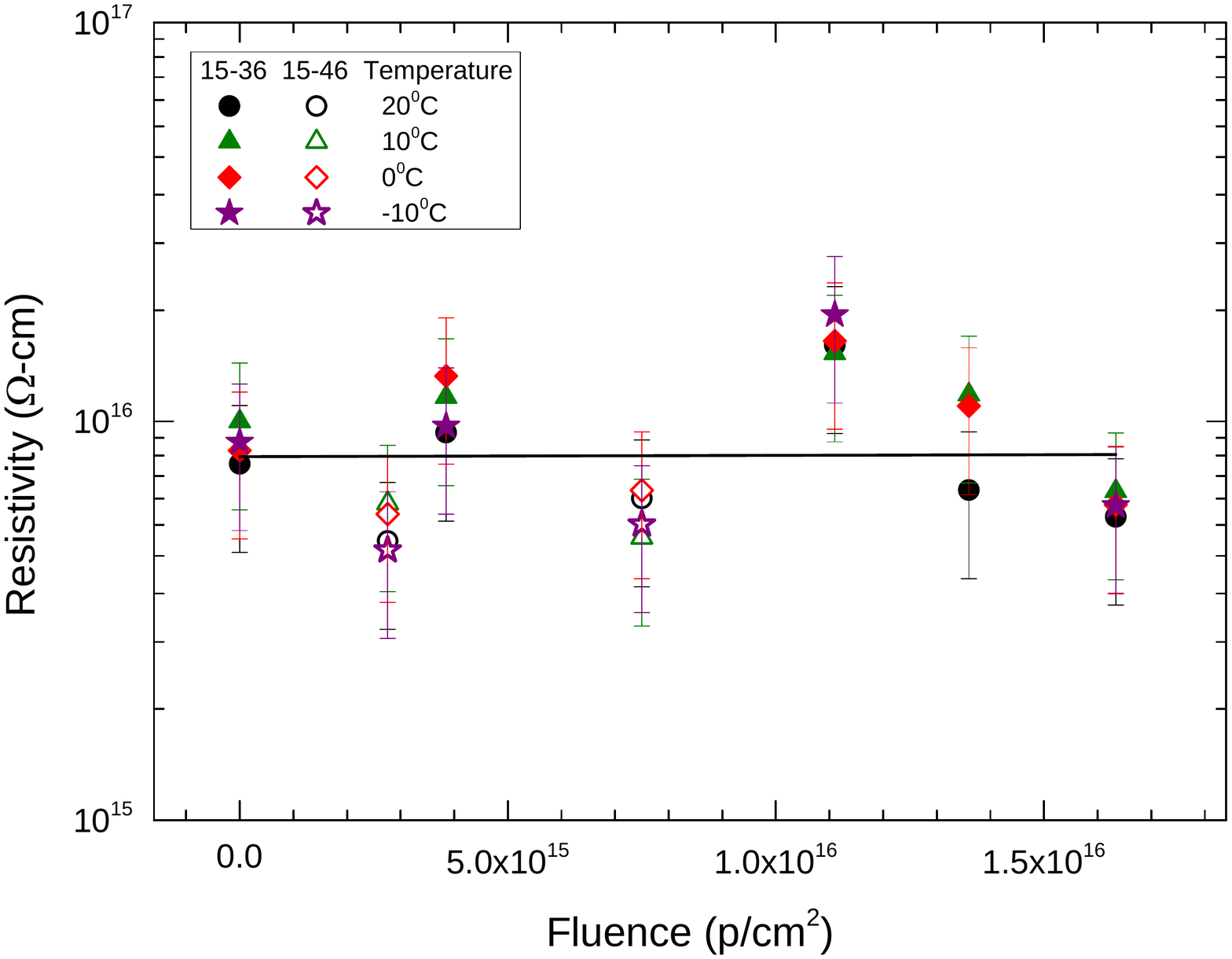}
        \caption{Resistivity of the diamond sensors as a function of fluence, for temperatures ranging from $-10^\circ$C to $+20^\circ$C.  A free linear fit to the data is shown, with intercept $(8.01 \pm 0.81) \times 10^{15} \Omega$-cm, slope
$(0.49 \pm 8.54) \times 10^{-2} \Omega$-cm/(p/cm$^2$), and $\chi^2$/dof 0.62.  The error bars show the combination of statistical
and systematic uncertainty.}
        \label{diamond-RF}
      \end{figure}

Figure~\ref{diamondrt} summarizes the resistivity versus temperature for all fluences
and both diamonds.  Figure~\ref{diamond-RF} summarizes the resistivity versus fluence for all temperatures and both diamonds.  The
error bars on these graphs are the quadrature sum of statistical and systematic uncertainties (see below).
A linear fit to the data in Figure~\ref{diamondrt} returns an intercept of
$(8.37 \pm 0.55) \times 10^{15} \Omega$-cm
and slope $(-0.63 \pm 4.13) \times 10^{13} \Omega$-cm/$^\circ$C, with $\chi^2$/dof 0.62.  A linear fit to the data in
Figure~\ref{diamond-RF} returns an intercept of $(8.01 \pm 0.81) \times 10^{15} \Omega$-cm and slope
$(0.49 \pm 8.4) \times 10^{-2} \Omega$-cm/(p/cm$^2$) with similar $\chi^2$/dof.
Thus diamond 1006115-36 and diamond 1006115-46 both have resistivity approximately $10^{16}$ Ohm-cm,
independent of fluence up to $1.63 \times 10^{16}$ 800-MeV p/cm$^2$ and independent of
temperature over the range [$-12^\circ$ C, $+23^\circ$ C].
The propagated uncertainties on the resistivity of the diamonds range
from 10 to 30\% for 1006115-36 and 10 to 40\% for 1006115-46.  Setup 2 was used for the measurements of
diamond 1006115-36 at fluence $1.63 \times 10^{16}$ p/cm$^2$ and temperatures $20^\circ$C, $0^\circ$C, and
$-10^\circ$C and for the measurements of diamond 1006115-46 at fluences $2.75 \times 10^{15}$ p/cm$^2$
and $7.5 \times 10^{15}$ p/cm$^2$ and temperatures $20^\circ$C, $10^\circ$C, $0^\circ$C, and
$-10^\circ$.  Setup 1 was used for the complement of the measurements.  Measurements of diamond 1006115-36 at
$20^\circ$C after fluence $1.63 \times 10^{16}$ p/cm$^2$ were made with both setups.

\subsection{Systematic Uncertainties}

The systematic uncertainties on the bias voltage and leakage current derive
from the manufacturer's accuracy specifications for the Keithley 237 and 617 and are
$\pm(0.04\% + 240~\rm{mV})$ and $\pm(0.3\%+100$ fA) respectively.  The uncertainties on
the
measured dimensions are given in the caption of Table~\ref{tab1}. The fluences are known to 10-30\%.  The systematic
uncertainty on the measured value of $\rho$ for each temperature and fluence condition
is obtained by shifting the measured voltages $\pm 1$ standard deviation while shifting
the measured currents $\mp 1$ standard deviation, then separately refitting the data in
Figure~\ref{fig:5ivs}.  All of these contributions yield a systematic uncertainty of magnitude
 1 to $6 \times 10^{14}$
Ohm-cm, about one order of magnitude smaller than the statistical uncertainties.
The systematic uncertainty associated with the setup configuration is 40\%.  This was determined
experimentally by measuring a device with both setups under the same conditions.  This
uncertainty is traceable to the fact that in the current and temperature regime
of this study, the intrinsic accuracy of the Keithley 617 (used
in Setup 2) is 1.6\% of the
reading whereas the intrinsic accuracy of the Keithley 237 (used in Setup 1) is 0.3\%.

\section{Conclusions}

The resistivity of chemical vapor deposition polycrystalline diamond sensors
has been measured for samples irradiated with 800 MeV protons in several steps
to a fluence of $1.63 \times 10^{16}$ p/cm$^2$.  The measurements were made
in a controlled-humidity environment over the temperature range approximately $-10^\circ$ to
$+20^\circ$ C.  We find no evidence for significant dependence of the resistivity upon either temperature
or particle fluence within the ranges studied.  The resistivity of two diamonds in the same series was
found to
be very consistent and of
order of magnitude of $10^{16}$ Ohm-cm.


\section{Acknowledgements}

We thank Harris Kagan for providing the diamond sensors.  This work was supported by
the U.S. Department of Energy.


\end{document}